# Dopant-vacancy binding effects in Li-doped magnesium hydride

Kyle C. Smith,[1] Timothy S. Fisher,[1] Umesh V. Waghmare[2] and Ricardo Grau-Crespo[3*]

[1] *School of Mechanical Engineering and Birck Nanotechnology Center, Purdue University, West Lafayette, IN 47907, USA*

[2] *Theoretical Sciences Unit, Jawaharlal Nehru Centre for Advanced Scientific Research, Jakkur Campus, Bangalore 560 064, India.*

[3] *Department of Chemistry, University College London, 20 Gordon Street, London WC1H 0AJ, UK.*

Corresponding author's email: r.grau-crespo@ucl.ac.uk

We use a combination of *ab initio* calculations and statistical mechanics to investigate the substitution of $Li^+$ for $Mg^{2+}$ in magnesium hydride ($MgH_2$) accompanied by the formation of hydrogen vacancies with positive charge (with respect to the original ion at the site). We show here that the binding energy between dopants and vacancy defects leads to a significant fraction of trapped vacancies and therefore a dramatic reduction of the number of free vacancies available for diffusion. The concentration of free vacancies initially increases with dopant concentration, but reaches a maximum at around 1 mol% Li doping and slowly decreases with further doping. At the optimal level of doping, the corresponding concentration of free vacancies is much higher than the equilibrium concentrations of charged and neutral vacancies in pure $MgH_2$ at typical hydrogen storage conditions. We also show that Li-doped $MgH_2$ is thermodynamically metastable with respect to phase separation into pure magnesium and lithium hydrides at any significant Li concentration, even after considering the stabilization provided by dopant-vacancy interactions and configurational entropic effects. Our results suggest that lithium doping may enhance hydrogen diffusion hydride, but only to a limited extent determined by an optimal dopant concentration and conditioned to the stability of the doped phase.

I.  INTRODUCTION

Although metal hydrides have the potential for reversible hydrogen storage and release at low temperatures and pressures on-board fuel cell vehicles, it has been difficult in practice to find combinations of metals with properties capable of meeting storage system performance targets. Magnesium hydride ($MgH_2$) possesses very high gravimetric and volumetric hydrogen storage capacity (7.6 wt % H and $6.5 \cdot 10^{22}$ H atoms/cm$^3$).[1] However, the release of hydrogen requires a phase transformation from $MgH_2$ (β phase) to hcp metal Mg with interstitial H (α phase), which occurs at temperatures too high (approximately 300°C in 1 bar $H_2$ gas) for practical operation.[2] Furthermore, the Mg-H system exhibits very slow kinetics of dehydrogenation and hydrogenation.[3] The kinetic behavior can be interpreted in terms of a model of phase nucleation and growth, where particles surfaces consist mainly of hydride (β phase) during hydrogenation and metallic Mg (α phase) during dehydrogenation.[4] The rate-limiting step for hydrogen absorption has been suggested to be the diffusion through the external hydride phase layer,[5] while in the dehydrogenation process the limiting step seems to be the propagation of the hydride-metal interface, which is controlled by the thermally activated emission of hydrogen across the interface.[6]

Despite these challenges significant interest exists in $MgH_2$ as a parent material for hydrogen storage applications, which is mainly driven by the attractively low molecular mass of magnesium. A well investigated route to improving the properties of the Mg-H system for hydrogen storage is the reduction of particle size, *e.g.* via ball milling. This type of mechanical treatment leads to faster kinetics due to shortening of diffusion length scales, increase of surface area and formation of diffusion-enhancing defects within the particles.[7] It has also been observed that $MgH_2$ nanoparticles exhibit improved thermodynamic behavior.  For example a decrease of up to 100°C in the onset temperature for desorption after ball-milling of $MgH_2$ samples was reported in Ref. [8]. An *ab initio* simulation study by Wagemans *et al.*[9] showed that structural rearrangements of $MgH_2$ clusters with sizes of ~1.3 nm and below lead to significant decreases of hydrogen desorption energy. It is therefore clearly desirable to minimize the size of the particles until reaching sub-nanometer length scales, but the upscaling of this strategy for commercial applications is not straightforward, as the process is energy intensive.

An alternative, or complementary, route for improving the kinetic performance of magnesium hydride in hydrogen storage is chemical doping. Several transition metal dopants have been investigated (*e.g.* Ti, V, Mn, Fe, and Ni in Ref. [10]), but doping with lighter elements (Li, B, N, *etc.*) would be preferable in order to retain the high gravimetric density of $MgH_2$. Johnson *et al.*[3] have doped $MgH_2$ with 10 wt % $LiBH_4$ and found the kinetics of hydrogen adsorption and desorption to increase significantly relative to pure $MgH_2$. The kinetic enhancement was observed only after 4-5 cycles of hydrogenation and dehydrogenation, when the sample was found to consist of B- and Li-doped rutile-type $MgH_2$, with no indication of borane ($[BH_4]^-$) anions. The exact roles of the B and Li dopants in the kinetic improvement were not clearly identified in this study. However, the observation that the use of $NaBH_4$ instead of $LiBH_4$ resulted in no kinetic improvement led the authors to suggest that the effect of Li doping could be significant, possibly by stabilizing hydrogen vacancies which would facilitate the diffusion of hydrogen in the hydride phase. Mao *et al.*[11] found similar enhancement with 10 and 25 wt % $LiBH_4$ composited with $MgH_2$. In this work, based on the observation by X-ray diffraction that no mixed Li-Mg phase was formed, a completely different mechanism for the kinetic enhancements was suggested, where Li acts as a catalyst in hydrogenation/dehydrogenation reactions.

The incorporation of Li into the $MgH_2$ lattice, which is the focus of the present work, is also relevant to the recently discovered application of Mg/LiH composites as negative electrode material for rechargeable lithium-ion batteries.[12] The performance of the electrode is attributed to the conversion reaction $MgH_2 + 2Li^+ + 2e^- \leftrightarrow Mg + 2LiH$, followed by an alloying process where Li ions react with magnesium to form $Li_xMg_{1-x}$.[13] The question of whether a mixed Li/Mg hydride phase is formed under the operating conditions of the battery is relevant to understanding the electrode behavior.

Mixed Li/Mg hydride phases have been investigated by other authors using both experimental and theoretical methods. Pfrommer *et al.*[14] hypothesized about the existence of a $LiMgH_3$ phase with perovskite structure, although their theoretical calculations showed that such a structure is energetically unstable with respect to decomposition into pure LiH and $MgH_2$ by approximately 60 kJ/mol. Ikeda *et al.*[15] synthesized $Li_xNa_{1-x}MgH_3$ by ball-milling of LiH, NaH, and $MgH_2$ with compositions ranging from 0 to 0.5 with Li substituting for Na, but at $x = 1$ a perovskite $LiMgH_3$ phase was not formed. Vajeeston[16] has recently suggested, based on *ab initio* calculations, that the ground state for the $LiMgH_3$ composition corresponds to a rhombohedral unit cell like that of the $LiTaO_3$ structure, although an ilmenite-type structure ($FeTiO_3$) is very close in energy (only ~1 kJ/mol of difference). Both structures are still slightly unstable with respect to decomposition into pure hydrides. The results from these studies suggest that in the Li-Mg-H system, lithium tends to separate as pure lithium hydride instead of forming a mixed Li/Mg hydride phase. However, the presence of lithium as a low concentration dopant in the rutile-like $MgH_2$ structure cannot be ruled out.

Recent theoretical work by Hao and Sholl examined the effect of different dopants (Li was not included) on the defect population in $MgH_2$ and how different types of defects enhance hydrogen diffusion within the hydride.[17,18] Some monovalent dopants such as Na promote the formation of hydrogen vacancies, which are positively charged relative to the original H$^-$ site. According to their results, these positively charged vacancies are not only easier to create but also diffuse more readily than neutral vacancies formed by partial reduction of the hydride. On the other hand, higher-valence dopants such as Co favor the formation of hydrogen interstitials, which they found to have the lowest diffusion barrier among hydrogen defects. This analysis is based on the calculation of the defect formation energies as function of the Fermi energy in the solid, which is altered by the presence of dopants, thus allowing an elegant formulation of the equilibrium between dopants and defects.[19] However, this approach does not include the treatment of direct interactions between dopants and defects, which can have a significant impact on diffusion. A well known example from the field of ionic conductors is the binding of oxygen vacancies to lower-valence dopants in fluorite oxides; this interaction is responsible for the decrease in ionic conductivity after reaching a maximum value with respect to doping concentration.[20]

In this paper we present the results of a theoretical investigation of the incorporation of Li$^+$ ions into the rutile-type $MgH_2$ lattice and the accompanying formation of one hydrogen vacancy per dopant. We give special attention to the interaction between dopants and vacancies, which we will show has a significant detrimental effect on the defect-mediated diffusion in the hydride. We also discuss the effect of these interactions on the thermodynamic stability of Li-doped $MgH_2$ with respect to phase separation.

II. METHODOLOGY

The relative distribution of dopants and hydrogen vacancies was investigated by generating all the symmetrically inequivalent configurations in supercells of different sizes, including 2x2x2, 2x2x3 and 3x3x4, each with one Li dopant and one H vacancy. In order to find the inequivalent configurations we follow the procedure implemented in the SOD (Site Occupancy Disorder) program, which is described in detail in Ref. [21], and has been previously employed in the description of species distribution over crystal sites in other materials (*e.g.* refs [22-25]). Within this approach, all possible configurations in a given supercell are initially generated, under the assumption that each configuration can be uniquely described by the enumeration of the substitution sites in a parent structure. After the full ensemble of configurations is generated, the inequivalent ones are selected and are fully relaxed. The criterion of equivalence between two configurations is the existence of an isometric transformation that converts one configuration into the other, where a list of possible symmetry operations is provided by the symmetry group of the the parent structure, combined with the supercell internal translations. This approach is correct as long as the symmetry of each configuration is preserved upon relaxation, as any symmetry breaking during relaxation would imply the existence of more configurations beyond those originally considered.

The energy of each configuration was then calculated using the density functional theory (DFT) with a generalized gradient approximation (GGA) functional built from the Perdew and Zunger[26] local functional and the gradient corrections by Perdew *et al.*,[27] using the VASP code.[28] The interaction between valence and core electrons was described with the projector augmented wave (PAW) method.[29] Core orbitals were kept frozen to the reference states, including levels up to $2p$ for Mg, and $1s$ for Li. Brillouin zone points were sampled using a Monkhorst-Pack mesh with density high enough to converge total energies within 1 meV per formula unit, leading to 2x2x2 partitions for the 2x2x3 and 3x3x4 supercells of $MgH_2$ and 4x4x4 for the cubic unit cell of LiH. A cutoff energy of 500 eV for the plane wave expansion, which provided a similar degree of convergence, was used here. Ionic positions were relaxed using conjugate gradient minimization until all the forces acting on ions were less than 0.01 eV/Å.

Once a complete spectrum was obtained *ab initio* for each supercell, a simpler model of the dopant-defect interaction was introduced in order to extrapolate the analysis to other dopant concentrations and examine the effect of the size of the simulation cell on our results. The details of this model are explained later when we discuss the results of our analysis using statistical mechanics.

## III. RESULTS AND DISCUSSION

### A. Configurational energy spectra and dopant – defect binding

We first discuss the total energy spectra of configurations with one Li substitution and one hydrogen vacancy in the 2x2x2, 2x2x3, and 3x3x4 supercells of the tetragonal $MgH_2$ unit cell. In the Kröger–Vink notation[30] this dopant-defect pair can be represented as $Li'_{Mg} + V^{\bullet}_{H}$, which emphasizes the negative and positive charges relative to the original lattice sites at dopant and vacancy sites, respectively. The total number of configurations for a supercell of composition $LiMg_{N-1}H_{2N-1}$ is in principle $2N^2$, which is excessive for accurate DFT calculations, but this number is drastically reduced when the symmetry of the lattice is taken into account, as shown in Table 1. We have checked that for all supercells the symmetry of each configuration is preserved upon relaxation. Table 2 lists the inequivalent configurations for the supercell 3x3x4 in ascending order of energy, together with their degeneracies, the relative positions of the dopant and the vacancy, the shortest distance $r_{min}$ between Li and H vacancy in the lattice (considering all the periodic images) and the symmetry space group.

The calculated configurational spectra for the three supercells are illustrated in Fig. 1. The energy range of each spectrum is significant, with energy differences of approximately 0.55 eV between the most and least stable configurations. Regardless of the supercell, the most stable configurations are those in which the vacancy is in a nearest neighbor (NN) position with respect to the dopant, *i.e.*, the vacancy is in the coordination octahedron of the Li ion. Because of the tetragonal symmetry of the crystal, two symmetrically inequivalent configurations of this type exist: one with two-fold degeneracy in which the $Li'_{Mg} + V^{\bullet}_H$ pair is orientated perpendicular to the [001] direction, and another with four-fold degeneracy where the defect pair forms an acute angle with the [001] direction. We find that the four-fold degenerate NN configuration is always the lowest in energy. In the smallest (2x2x2) supercell, where dopants are only about 6 Å apart along the [001] direction, the gap between the four-fold and two-fold NN configuration is substantially larger than in the larger supercells; this effect is due to the closer interaction between the periodic images of the defects in that cell. As shown in Fig. 1, the energy gap between inequivalent NN configurations reduces considerably for the 2x2x3 supercell but is practically converged for the 3x3x4 supercell. The remaining anisotropy reflected in the energy splitting even for the largest supercell is consistent with the tetragonal symmetry of the lattice.

In the high-energy end of the spectra, the density of configurations increases with supercell size, and for the largest supercell (3x3x4) most configurations are accumulated at the top of the spectrum. These configurations correspond to larger distances and therefore weaker interactions between dopant and defect. The configuration energies for the 3x3x4 supercell are plotted against the dopant-defect separation in Fig. 2; the separation $r_{min}$ is the minimum of the distances between Li ions and H vacancies in the lattice, considering all the periodic images. The shortest distance of about 2 Å corresponds to the two NN configurations, and the energy increases with separation, reaching a maximum for the longest possible dopant-defect distance in this supercell, at about 9 Å separation. We observe a clear separation in energy of NN configurations from the remaining configurations.

Notably, the energy difference of 0.55 eV between the lowest- and highest-energy configurations is much less than the value expected from pure electrostatics: a 5.2 eV difference in Coulomb energy was calculated with the GULP code,[31] assuming formal charges for the defect species and using Ewald summations to include interactions with images. This behavior does not seem to be caused by any strong departure from the ionic character of the interactions, as we have checked that the electronic charge in the vicinity of the Li$^+$ dopant does not change significantly in the presence of a NN hydrogen vacancy. Of course, the polarization of the ions introduces some deviation from perfect ionicity. An additional contribution to the reduction of interaction energy results from the local distortion of the lattice required to accommodate the dopant-defect pair in NN configurations, and can be rationalized in terms of ionic radii as follows. The size of Li$^+$ (0.76 Å) is slightly larger than that of Mg$^{2+}$ (0.72 Å),[32] and as a result, in the absence of H vacancies in the first coordination shell, Li-H distances (2 x 1.96 Å, 4 x 1.99 Å, see Fig. 3a) are somewhat elongated with respect to Mg-H distances in the bulk hydride (2 x 1.94 Å, 4 x 1.95 Å). If a hydrogen vacancy is created around a Li cation, the other five hydrogen ions in the coordination octahedron move closer to Li$^+$ (Fig. 3b). In particular, the hydrogen ion opposite to the hydrogen vacancy shifts its position significantly towards the cation (Li-H distance of 1.80 Å in Fig. 3b), due to the force imbalance resulting from the absence of the negative ion on the opposite side. Although a similar distortion occurs around each Mg$^{2+}$ center neighboring a hydrogen vacancy, we can expect that the energetic effect of the distortion introduced by the vacancy will be stronger for the NN configurations than for the others, because the larger size of the Li$^+$ cation results in the shrinking distortion of the coordination octahedron being energetically more expensive for Li$^+$ than for Mg$^{2+}$. The geometric distortion created by the vacancy around Li is significant, so we predict that the presence of $Li'_{Mg} + V^{\bullet}_H$ pairs in NN configuration could be easily detected from EXAFS measurements of Li-doped MgH$_2$.

## B. Statistical mechanical analysis

Having established the energetic preference for hydrogen vacancies to locate in NN positions with respect to the Li dopant, the question remains of what fraction of vacancies are trapped by dopants when the material is in thermodynamic equilibrium. This question is relevant for hydrogen storage applications, as dopant-defect binding inhibits the ability of hydrogen vacancies to diffuse within the hydride. We expect that the number of free vacancies (defined here as those not NN to the dopants) will increase with temperature at a given dopant concentration. At fixed dopant concentration $x$ and number of MgH$_2$ formula units $N$, we can assign a Boltzmann probability to each configuration:[22,23,33]

$$P_m = \frac{\Omega_m}{Z} \exp(-E_m / k_B T), \qquad (1)$$

where $m=1, \ldots, M$ is the index of the configuration ($M$ is the number of inequivalent configurations), $E_m$ is the energy of the configuration, $\Omega_m$ is its degeneracy (the number of times that configuration $m$ is repeated in the complete configurational space), $k_B$ is Boltzmann's constant, and

$$Z = \sum_{m=1}^{M} \Omega_m \exp(-E_m / k_B T) \qquad (2)$$

is the canonical partition function. The concentration (number per $Li_xMg_{1-x}H_{2-x}$ formula unit) of free vacancies can then be calculated as:

$$x_{free} = x \sum_{m \neq NN} P_m \qquad (3)$$

where the summation is over all non-NN configurations.

Fig. 4 shows the calculated concentrations of free vacancies as obtained for the 3x3x4 ($x=0.014$) and 2x2x3 ($x=0.042$) supercells. In both cases, the number of free vacancies in the equilibrium distribution at $T=300$ K is insignificant compared with the total concentration of vacancies, *i.e.*, most vacancies are trapped by the dopant. The number of free vacancies increases quite rapidly with temperature, but even at 800 K these mobile vacancies represent only a small fraction of the total vacancy population. Higher temperatures are not considered here because the hydride phase would not be stable at those temperatures for typical partial pressures of hydrogen in storage applications. Clearly, our results suggest a very significant effect of dopant-defect interaction on the ability of the hydrogen vacancies to diffuse in the hydride.

Although the foregoing analysis allows an estimation of the concentration of free vacancies ($x_{free}$) and its dependence on temperature, it is not adequate for the discussion of the variation of $x_{free}$ with the concentration $x$ of dopants, which is of more practical interest. The reason for such inadequacy is that at each concentration we have employed a *minimal* supercell, *i.e.*, a cell that yields the desired composition when containing only one dopant and one vacancy. Although this simplification of the configurational space is convenient to perform accurate DFT calculations, it is insufficient for a proper statistical mechanical analysis as is shown below, if one intends to compare results at different dopant concentrations. The correct procedure is to employ a very large supercell, where all concentrations under study can be achieved by varying the number of dopants (and vacancies); only in this way are the configurational spaces at different concentrations comparable. Obviously, such an approach is beyond the reach of DFT calculations, and even of less accurate atomistic simulation methods, not only because of the large supercell required for describing each configuration, but also because of the vast number of possible configurations of dopants and vacancies.

It is possible, however, to achieve an exact treatment of configurational counting for very large supercells if the interaction between dopant and defect is treated in a manner that retains the trapping effect while approximating the energy spectrum more simply. In such a model the energy of interaction between dopants and defects in NN configurations is $-\varepsilon$, while the interaction is negligible for other configurations. The number of distinct levels in the configurational spectrum then reduces to $n+1$, where $n$ is the number of dopants (or vacancies) in the supercell; the energy is minimum when all vacancies are trapped and maximum when all vacancies are free. The configurational energies within this simplified model can be expressed in the following way. Each configuration in a supercell of $N$ formula units, with $n=xN$ dopants and the same number of vacancies, $t$ of which are trapped in NN positions, has the energy:

$$E_t = N(E_{MgH_2} + x\Delta E) - t\varepsilon \qquad (4)$$

where $E_{MgH_2}$ is the DFT energy per formula unit of pure MgH$_2$ and $\Delta E$ is the energy introduced by the substitution of one dopant and one hydrogen vacancy with no interaction between them. For fixed $N$ and $x$, each configurational level has an energy that depends only on $t$, and a degeneracy that is easily calculated using simple combinatorial analysis, even for large $N$. We have not considered here the effect of dopant-dopant and vacancy-vacancy repulsive interactions, which can be explained as follows. The relative dopant-vacancy distribution is determined by a competition between energetic stabilization upon pairing (proportional to the binding energy) and the entropic tendency to homogenization that results from the higher degeneracy of disassociated states (proportional to temperature). However, in the case of dopant-dopant and vacancy-vacancy distributions, both energetic and entropic effects tend to keep the defect species as far from each other as possible; thus at the low dopant concentrations of interest in this work (1-4%), with large average Li – Li distances, the effective contribution of these interactions to the configuration spectrum will be significantly less important than the contribution from dopant-vacancy interactions.

We utilize the full configurational spectrum obtained by DFT to determine parameters in the simplified interaction model. In particular, $\Delta E$ is calculated from the DFT energy of the configuration with the maximum vacancy-defect separation in the largest supercell. The dopant-defect binding energy $\varepsilon$ for our interaction model can be approximated by the difference in energy between NN configuration and the configuration having largest separation between dopant and vacancy in the largest supercell, which corresponds to the width of the DFT configurational spectrum (~0.55 eV). However, we have used a slightly more sophisticated approach to compensate somewhat for the drastic simplification of the configurational spectrum and the finite size supercell; the value of $\varepsilon$ was optimized to minimize the difference in configurational entropy[21,23]

$$S = k_B \sum_m P_m \ln \frac{\Omega_m}{P_m} \qquad (5)$$

between the full DFT spectrum in the minimal supercell $LiMg_{71}H_{143}$ and the simplified two-level spectrum in the same supercell. We note that the definition of 'configuration' is not the same in the two cases, and for the statistics in the simplified model the sum in Eq. (5) must be performed over all the values of $t$. This procedure yielded an effective binding energy $\varepsilon=0.475$ eV, which is slightly less than the width of the DFT spectrum, as expected from the fact that dopant-defect interactions are neglected beyond NN distances. By following this approach the simplified interaction model has been made thermodynamically consistent with the DFT supercell calculations at the minimal supercell, and can now be extended to larger supercells. We note that in our statistical treatment with the simplified interaction model at larger supercells, no approximation is made in terms of the size of the configurational space, which is always treated exactly, but only in the energy calculation.

We now discuss the selection of an adequate supercell size. Fig. 5 shows the temperature dependence of the configurational entropy of $Li_xMg_{1-x}H_{2-x}$, as calculated from the simplified interaction model for increasing supercell sizes at $T=600$ K. Since entropy changes very rapidly with $x$, in Fig. 5 it is plotted normalized by the limit entropy $S_{noint}$ in the absence of dopant-vacancy interaction (or equivalently at infinite temperature) at the given composition for an infinite supercell size. $S_{noint}$ can be obtained analytically as:

$$S_{noint} = -k_B \left( x \ln x + (1-x) \ln(1-x) + x \ln(x/2) + (2-x) \ln(1-x/2) \right) \quad (6)$$

Also plotted in Fig. 5 is the maximum entropy of the simplified interaction model (in the absence of dopant-vacancy interaction or at infinite temperature) as calculated for each supercell size; convergence of this normalized entropy to unity in the limit of infinite supercell size reflects consistency of the simplified interaction model. These plots clearly demonstrate that adequate convergence can be achieved for $N=1000$, because for this supercell size the error of the maximum entropy is less than 4% at worst for the concentration range presented. In what follows we have used this supercell size for our analysis of trapping and stability in the doped hydride.

The concentration of free vacancies can be now calculated from the Boltzmann's probabilities $P_t$ associated with configurational levels $t=0$ to $xN$:

$$x_{free} = x - \sum_{t=0}^{xN} tP_t, \qquad (7)$$

and the result is shown as a function of $x$ in Fig. 6. As discussed previously regarding the statistics in the minimal supercell, only a small fraction of vacancies is free from dopants, and this fraction increases with temperature. However, contrasting with the results from the minimal supercell analysis, the concentration of free vacancies at 4% Li doping is not higher than at 1% doping. In fact, $x_{free}$ increases with Li concentration up to ~1% and then slightly decreases at higher dopant concentrations. This difference illustrates the necessity of using a fixed (and large) supercell size for the statistical mechanical analysis at variable concentration. More importantly, our results indicate that, although Li doping is an efficient way of introducing hydrogen vacancies in the hydride, doping levels beyond ~1% can detrimentally affect vacancy diffusivity in the material, as the higher abundance of trapping centers leads to a decrease in the concentration of free vacancies relative to that at the optimal doping level. The maximum concentration of free vacancies ranges from $4 \cdot 10^{-5}$ to $1.5 \cdot 10^{-4}$ at 600 to 700 K, implying that at least 98% of the total vacancies in the material are trapped. Despite this strong trapping effect, the concentration of free vacancies generated by optimally doping with Li is substantially larger than both our previous prediction for the fraction of neutral vacancies in pure $MgH_2$ ($10^{-7}$ to $10^{-6}$)[24] and Hao and Sholl's prediction for the concentration of charged vacancy-interstitial pairs in pure $MgH_2$ ($10^{-11}$ to $10^{-10}$).[17] Therefore, despite strong vacancy trapping by Li doping, this result suggests that the presence of Li dopants could significantly enhance total vacancy diffusivity in $MgH_2$.

## C. Mixing thermodynamics

We now examine the stability of Li-doped $MgH_2$ with respect to hydride phases of the constituent metals, LiH and $MgH_2$. We define the mixing enthalpy $H_{mix}$ as:

$$H_{mix}(x,T) = H(x,T) - (1-x)E_{MgH_2} - xE_{LiH} \qquad (7)$$

where $E_{MgH_2}$ and $E_{LiH}$ are the DFT bulk energies per formula unit of pure $MgH_2$ and LiH, and

$$H(x,T) = \frac{1}{N}\sum_{t=0}^{xN} P_t E_t \qquad (8)$$

is the configurational average of energies (per formula unit). The enthalpy of mixing is strongly positive for the mixed hydride over the composition range investigated here (0.01<$x$<0.04). Since our calculations already correspond to very low dopant concentrations, the sign of the mixing enthalpy is expected to remain positive even in the limit of infinite dilution, and we can define the effective formation enthalpy of the dilute dopant-defect pair as:

$$H_f(T) = \lim_{x \to 0} \frac{H_{mix}(x,T)}{x}, \qquad (9)$$

which we evaluate using a very small value of $x$. In the absence of dopant-vacancy interaction (or equivalently, in the limit of infinite temperature), the formation enthalpy of the $Li'_{Mg} + V^{\bullet}_H$ pair is 1.55 eV. Because of dopant-defect interactions, the formation energy decreases to 1.08 eV, with no significant variation for temperatures between 300 and 600 K.

However, the reduction in the formation energy due to dopant-defect interaction is not enough to stabilize any significant concentration of Li in a mixed hydride with respect to phase separation into the pure hydrides. This conclusion can be obtained from the analysis of the mixing

free energy:

$$G_{mix}(x,T) = G(x,T) - (1-x)E_{MgH_2} - xE_{LiH} \quad (10)$$

where

$$G(x,T) = -\frac{1}{N}k_B T \ln Z \quad (11)$$

is the configurational free energy per formula unit of the mixed hydride. The configurational entropy $S(x,T)$ is included in the free energy but not in the enthalpy, and therefore:

$$S(x,T) = \frac{H(x,T) - G(x,T)}{T} \quad (12)$$

The entropic effect tends to stabilize mixing, but the resulting mixing free energy is still positive for the range of concentrations considered here (0.01<$x$<0.04). The mixing free energy would only become negative at extremely low dopant concentrations, of the order of $\exp(-H_f/k_B T)$ (*e.g.* ~$10^{-9}$ at $T$=600 K) or less. Therefore, we can conclude that doping MgH$_2$ with any significant concentration of Li is thermodynamically unstable with respect to phase separation, even after considering the stabilization provided by dopant-vacancy interactions and configurational entropic effects.

It should not be inferred from the above conclusion that it is impossible in practice to dope Li into magnesium hydride. In fact, the decomposition of the mixed phase into pure hydrides, although thermodynamically favorable, is kinetically forbidden at the conditions of interest here. The separation of phases requires cation diffusion, which involves high energetic barriers within the solid phase. Therefore, at the relatively low temperatures of interest in hydrogen storage (below 600 K) bulk phase separation is not expected to occur at a significant rate. Under these conditions, we can refer to Li-doped magnesium hydride as being *metastable* with respect to phase separation. Metastable mixed phases can be routinely prepared with an adequate choice of synthesis procedure. A well known example is the $Ce_{1-x}Zr_xO_2$ solid solution, which despite being metastable with respect to the pure oxides,[34] can be synthesized in the whole composition range and it is commonly used as part of the catalyst in exhaust catalytic converters.[35] The experimental study of Johnson *et al.*,[3] involving the incorporation of Li and B into magnesium hydride via reactive ball-milling, has found the presence of only $MgH_2$-like rutile phase in the mixed hydride using X-ray diffraction, thus confirming the absence of phase separation. Nevertheless, our results indicate that the stability of Li-doped magnesium hydride phases, even at low Li concentration, relies only on kinetic barriers preventing decomposition, and therefore solid-state material processing must be tailored to prevent separation of the doped phase at moderately high temperatures and/or long operation times.

## IV. SUMMARY

Monovalent dopants like Li can stabilize positive hydrogen vacancies in magnesium hydride, with the potential to improve hydrogen diffusion. However, our calculations show that dopant-defect attractive interactions are likely to have a significant detrimental role in the diffusion properties of the hydride. The binding energy between a Li dopant and a hydrogen vacancy nearest to the dopant is approximately 0.5 eV, which is enough to immobilize most (at least 98%) of the hydrogen vacancies at temperatures of interest in hydrogen storage applications.

At moderately low dopant concentrations (~1%) a maximum in free vacancy concentration is observed. At this optimal doping level, the concentration of free vacancies, although small in comparison with the total concentration of vacancies, is already two orders of magnitude higher than the concentration of neutral and charged vacancies previously predicted for undoped $MgH_2$. Therefore, doping with Li should still enhance vacancy mediated-diffusion in $MgH_2$. The magnitude of this enhancement cannot be evaluated from the present investigation, as we have not calculated diffusion barriers. However, previous results by other authors indicate that positive hydrogen vacancies do offer a lower barrier for diffusion than neutral vacancies, although other mechanisms of diffusion involving hydrogen interstitials may exhibit even lower diffusion barriers.[17,18] Finally, we have shown that Li-doped $MgH_2$ is metastable with respect to decomposition into pure Li and Mg hydrides, and this finding should influence the techniques employed to synthesize Li-doped $MgH_2$. In conclusion, the present results, and particularly the quantification of vacancy trapping effects, provide a better atomic-level understanding of the difficulties involved in the improvement of the diffusion kinetics of light-metal hydrides for hydrogen storage applications.

**ACKNOWLEDGEMENTS**

This work made use of the facilities of HECToR, the UK's national high-performance computing service, via RGC's membership of the UK's HPC Materials Chemistry Consortium, which is funded by EPSRC (EP/F067496). KCS acknowledges the Indo-US Science and Technology Forum for supporting his visit to JNCASR and UCL through the Joint Networked Centre on Nanomaterials for Energy (Award 115-2008/2009-10) and thanks the US National Science Foundation for financial support via a graduate research fellowship. UVW thanks the Indian Department of Atomic Energy for an Outstanding Researcher Fellowship.

**Table 1**

**Total number of configurations ($M_{total}$) and number of symmetrically inequivalent configurations ($M$) calculated with DFT for each composition.**

| Supercell | Composition | $M_{total}$ | $M$ |
|---|---|---|---|
| 2x2x2 | LiMg$_{15}$H$_{31}$ | 512 | 9 |
| 2x2x3 | LiMg$_{23}$H$_{47}$ | 1152 | 12 |
| 3x3x4 | LiMg$_{71}$H$_{143}$ | 10368 | 30 |

**Table 2**

Inequivalent configurations of one Li dopant and one H vacancy in the 3x3x4 supercell of $MgH_2$. $E$ is the energy relative to the most stable configuration, $\Omega$ is the degeneracy, $(x, y, z)$ is the position of the H vacancy in fractional coordinates, and $r_{min}$ is the shortest distance between the dopant and the vacancy. The Li dopant is always positioned at (0,0,0).

| $E$ (eV) | $\Omega$ | $x$ | $y$ | $z$ | $r_{min}$ (Å) | Space group |
|---|---|---|---|---|---|---|
| 0.000 | 288 | 0.065 | 0.935 | 0.125 | 1.95 | Cm |
| 0.046 | 144 | 0.102 | 0.102 | 0.000 | 1.94 | Amm2 |
| 0.292 | 144 | 0.768 | 0.768 | 0.000 | 4.42 | Amm2 |
| 0.379 | 288 | 0.102 | 0.102 | 0.250 | 3.57 | Cm |
| 0.387 | 288 | 0.065 | 0.935 | 0.375 | 4.67 | Cm |
| 0.398 | 288 | 0.768 | 0.768 | 0.250 | 5.35 | Cm |
| 0.407 | 288 | 0.102 | 0.768 | 0.000 | 3.42 | Pm |
| 0.427 | 288 | 0.732 | 0.268 | 0.125 | 5.33 | Cm |
| 0.438 | 576 | 0.102 | 0.768 | 0.250 | 4.55 | P1 |
| 0.460 | 576 | 0.065 | 0.268 | 0.125 | 4.02 | P1 |
| 0.466 | 144 | 0.102 | 0.102 | 0.500 | 6.31 | Amm2 |
| 0.474 | 576 | 0.065 | 0.602 | 0.125 | 5.65 | P1 |
| 0.475 | 288 | 0.398 | 0.602 | 0.125 | 7.75 | Cm |
| 0.488 | 288 | 0.732 | 0.268 | 0.375 | 6.82 | Cm |
| 0.495 | 288 | 0.102 | 0.435 | 0.000 | 6.03 | Pm |
| 0.498 | 144 | 0.435 | 0.435 | 0.000 | 8.30 | Amm2 |
| 0.508 | 288 | 0.435 | 0.768 | 0.000 | 6.65 | Pm |
| 0.510 | 288 | 0.435 | 0.435 | 0.250 | 8.83 | Cm |
| 0.510 | 576 | 0.398 | 0.268 | 0.125 | 6.65 | P1 |
| 0.514 | 288 | 0.102 | 0.768 | 0.500 | 6.91 | Pm |
| 0.514 | 576 | 0.065 | 0.268 | 0.375 | 5.84 | P1 |
| 0.515 | 144 | 0.768 | 0.768 | 0.500 | 7.46 | Amm2 |
| 0.518 | 288 | 0.398 | 0.602 | 0.375 | 8.84 | Cm |
| 0.519 | 576 | 0.102 | 0.435 | 0.250 | 6.73 | P1 |
| 0.526 | 576 | 0.435 | 0.768 | 0.250 | 7.30 | P1 |
| 0.526 | 576 | 0.065 | 0.602 | 0.375 | 7.07 | P1 |
| 0.531 | 288 | 0.102 | 0.435 | 0.500 | 8.51 | Pm |
| 0.541 | 576 | 0.398 | 0.268 | 0.375 | 7.89 | P1 |
| 0.543 | 288 | 0.435 | 0.768 | 0.500 | 8.96 | Pm |
| 0.545 | 144 | 0.435 | 0.435 | 0.500 | 10.24 | Amm2 |

**Figure captions**

**Figure 1.** Configurational energy spectrum of Li-doped $MgH_2$ for various supercell sizes. The dashed box encloses the configurations with vacancies in nearest neighbor (NN) positions with respect to dopants.

**Figure 2.** Configurational energy as a function of minimum distance between the Li dopant and the H vacancy in doped $MgH_2$ for a 3x3x4 supercell, with composition $LiMg_{24}H_{47}$.

**Figure 3.** Coordination octahedron of a Li dopant in $MgH_2$ (a) in the absence and (b) in the presence of nearest neighbor (NN) hydrogen vacancy.

**Figure 4.** Concentration ($x_{free}$) of free vacancies as a function of temperature as calculated with the minimum supercell size for each concentration (3x3x4 for $x=1/72$ and 2x2x3 for $x=1/24$).

**Figure 5.** Convergence of configurational entropy with supercell size for different dopant concentrations, at $T = 600$ K (curves at the bottom) and in the limit $\varepsilon/k_BT \rightarrow 0$ (curves at the top).

**Figure 6.** Concentration ($x_{free}$) of free vacancies as a function of the concentration of Li dopant in $MgH_2$, as calculated from the two-level trapping model for a large supercell ($N = 1000$).

**Figure 1.**

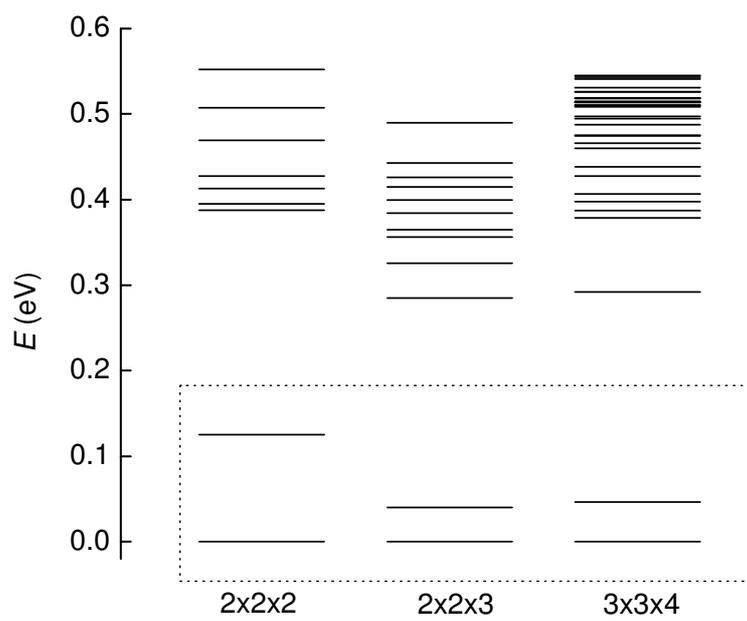

**Figure 2**

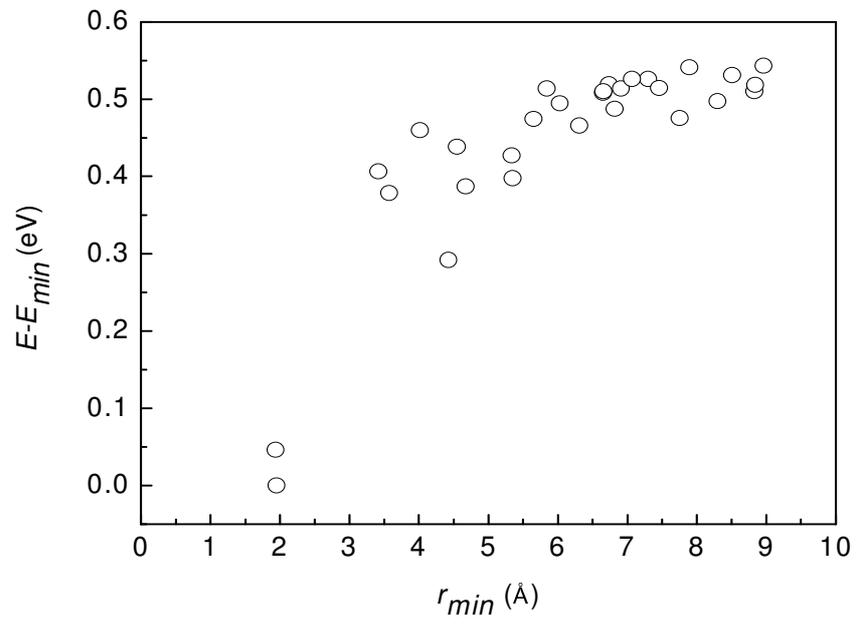

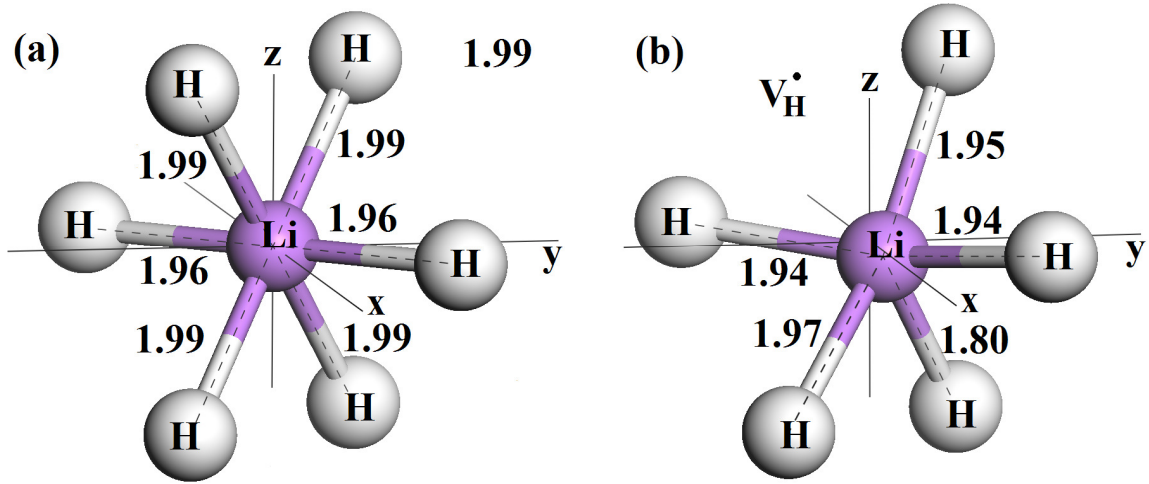

**Figure 4**

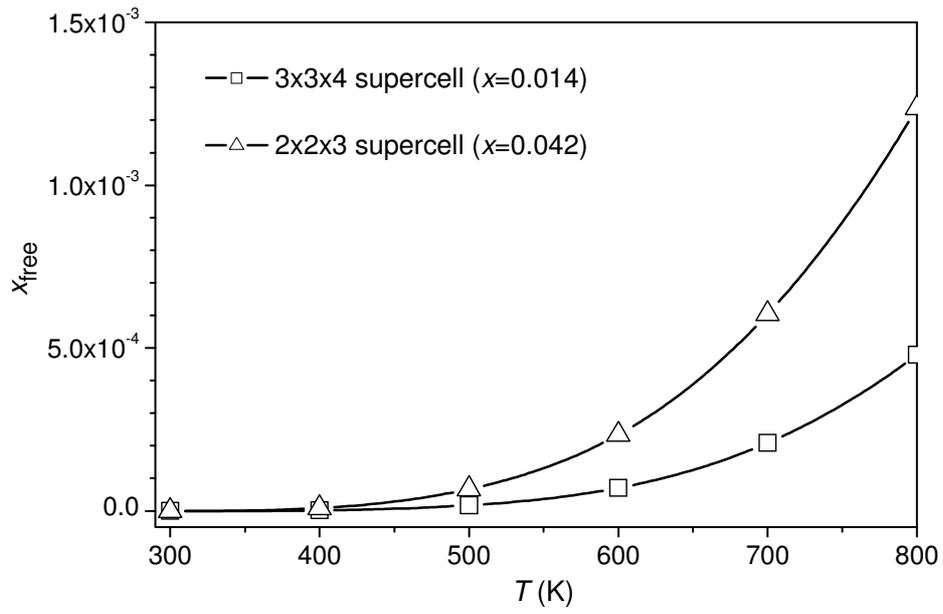

**Figure 5**

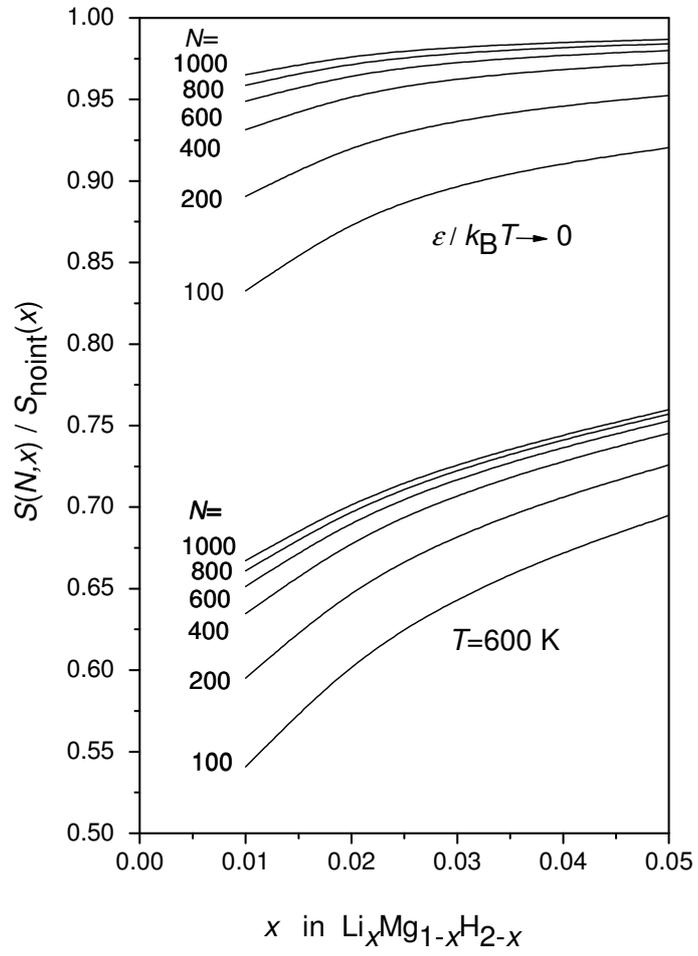

**Figure 6**

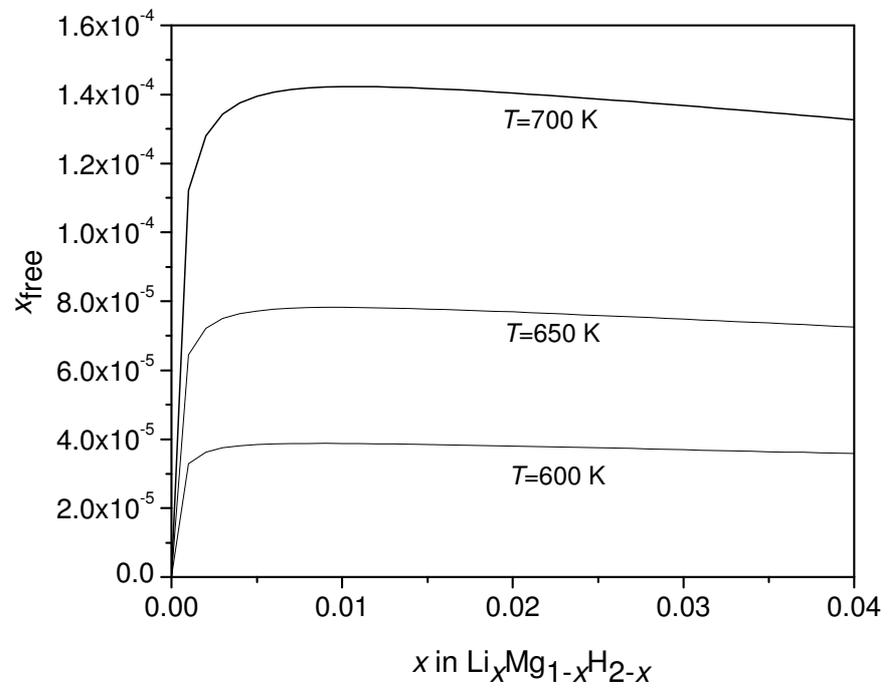

# REFERENCES

1. P. Selvam, B. Viswanathan, C. S. Swamy, and V. Srinivasan, Int. J. Hydrogen Energ. **11**, 169 (1986).

2. J. F. Stampfer, C. E. Holley, and J. F. Suttle, J. Am. Chem. Soc. **82**, 3504 (1960).

3. S. R. Johnson, P. A. Anderson, P. P. Edwards, I. Gameson, J. W. Prendergast, M. Al-Mamouri, D. Book, I. R. Harris, J. D. Speight, and A. Walton, Chem. Commun., 2823 (2005).

4. P. S. Rudman, J. Appl. Phys. **7195** 7195 (1979); M. V. C. Sastri, B. Viswanathan, and R. S. Achyuthlal Babu, in *Metal Hydrides: Fundamentals and Applications* edited by M. V. C. Sastri (Springer-Verlag New York, 1998).

5. B. Vigeholm, J. Kjoller, B. Larsen, and A. S. Pedersen, Journal of the Less-Common Metals **89**, 135 (1983); G. Friedlmeier and M. Groll, J. Alloy. Compd. **253-254**, 550 (1997).

6. G. Liang, J. Huot, S. Boily, and R. Schulz, J. Alloy. Compd. **305**, 239 (2000).

7. B. Sakintuna, F. Lamari-Darkrim, and M. Hirscher, Int. J. Hydrogen Energ. **32**, 1121 (2007).

8. A. Zaluska, L. Zaluski, and J. O. Ström-Olsen, J. Alloy. Compd. **289**, 197 (1999).

9. R. W. P. Wagemans, J. H. van Lenthe, P. E. de Jongh, A. J. van Dillen, and K. P. de Jong, J. Am. Chem. Soc. **127**, 16675 (2005).

10. G. Liang, J. Huot, S. Boily, A. Van Neste, and R. Schulz, J. Alloy. Compd. **292**, 247 (1999).

[11] J. F. Mao, Z. Wu, T. J. Chen, B. C. Weng, N. X. Xu, T. S. Huang, Z. P. Guo, H. K. Liu, D. M. Grant, G. S. Walker, and X. B. Yu, J. Phys. Chem. C **111**, 12495 (2007).

[12] Y. Oumellal, A. Rougier, G. A. Nazri, J. M. Tarascon, and L. Aymard, Nature Materials **7**, 916 (2008).

[13] Y. Oumellal, A. Rougier, J. M. Tarascon, and L. Aymard, J. Power Sources **192**, 698 (2009).

[14] B. Pfrommer, C. Elsasser, and M. Fahnle, Phys. Rev. B **50**, 5089 (1994).

[15] K. Ikeda, Y. Nakamori, and S. Orimo, Acta Materialia **53**, 3453 (2005).

[16] P. Vajeeston, P. Ravindran, A. Kjekshus, and H. Fjellvag, J. Alloy. Compd. **450**, 327 (2008).

[17] S. Q. Hao and D. S. Sholl, Appl. Phys. Lett. **93** (2008).

[18] S. Q. Hao and D. S. Sholl, Appl. Phys. Lett. **94** (2009).

[19] C. G. Van de Walle and J. Neugebauer, J. Appl. Phys. **95**, 3851 (2004).

[20] J. A. Kilner and B. C. H. Steele, in *Non-stoichiometric oxides*, edited by O. T. Sorensen (Academic Press, New York, 1981); C. R. A. Catlow, Solid State Ion. **12**, 67 (1984).

[21] R. Grau-Crespo, S. Hamad, C. R. A. Catlow, and N. H. de Leeuw, J. Phys. - Condens. Mat. **19**, 256201 (2007).

[22] R. Grau-Crespo, N. H. de Leeuw, and C. R. A. Catlow, Chem. Mater. **16**, 1954 (2004); S. E. Ruiz-Hernandez, R. Grau-Crespo, A. R. Ruiz-Salvador, and N. H. De Leeuw, Geochim. Cosmochim. Acta **74**, 1320 (2010).

[23] S. Benny, R. Grau-Crespo, and N. H. De Leeuw, Phys. Chem. Chem. Phys. **11**, 808 (2009).

[24] R. Grau-Crespo, K. C. Smith, T. S. Fisher, N. H. De Leeuw, and U. V. Waghmare, Phys. Rev. B **80**, 174117 (2009).

25  R. Grau-Crespo, A. Y. Al-Baitai, I. Saadoune, and N. H. De Leeuw, J. Phys.- Condens. Mat. **22**, 255401 (2010).

26  J. P. Perdew and A. Zunger, Phys. Rev. B **23**, 5048 (1981).

27  J. P. Perdew, J. A. Chevary, S. H. Vosko, K. A. Jackson, M. R. Pederson, D. J. Singh, and C. Fiolhais, Phys. Rev. B **46**, 6671 (1992).

28  G. Kresse and J. Furthmuller, Comput. Mater. Sci. **6**, 15 (1996); G. Kresse and J. Furthmuller, Phys. Rev. B **54**, 11169 (1996); G. Kresse and J. Hafner, Phys. Rev. B **48**, 13115 (1993); G. Kresse and J. Hafner, J. Phys. - Condens. Mat. **6**, 8245 (1994).

29  P. E. Blochl, Phys. Rev. B **50**, 17953 (1994); G. Kresse and D. Joubert, Phys. Rev. B **59**, 1758 (1999).

30  F. A. Kroger and H. J. Vink, in *Solid State Physics*, edited by F. Seitz and D. Turnbull (Academic Press, New York, 1956), Vol. 3, pp. 307.

31  J. D. Gale and A. L. Rohl, Mol. Simul. **29**, 291 (2003).

32  R. D. Shannon, Acta Crystallographica **A32**, 751 (1976).

33  R. Grau-Crespo, A. G. Peralta, A. R. Ruiz-Salvador, A. Gomez, and R. Lopez-Cordero, Phys. Chem. Chem. Phys. **2**, 5716 (2000).

34  T. A. Lee, C. R. Stanek, K. J. McClellan, J. N. Mitchell, and A. Navrotsky, Journal of Materials Research **23**, 1105 (2008).

35  R. Di Monte and J. Kaspar, Catal. Today **100**, 27 (2005).